\RequirePackage{ifpdf,xcolor,tikz}
\documentclass[hyper,letterpaper,notoc]{article}
\usepackage{jheppub}
\usepackage[utf8x]{inputenc}
\usepackage{amsfonts}
\usepackage{amsmath,amssymb,amscd}
\usepackage{xcolor}
\usepackage{array}
\usepackage{bbold}
\usepackage{tikz}
\usepackage{ytableau}
\def\bydef{\stackrel{\mathrm{def}}{=}}
\usepackage[numbers]{natbib}
\def\Tr{\mathrm{Tr}}

\newcommand{\twoh}
{\begin{tikzpicture}
		\draw (0.4,0) rectangle (0.8,0.2);
		\draw (0.6,0)--(0.6,0.2);
\end{tikzpicture}}
\newcommand{\twov}
{\begin{tikzpicture}
		\draw (0.4,0.2) rectangle (0.6,-0.2);
		\draw (0.4,0)--(0.6,0);
\end{tikzpicture}}

\title{Proving superintegrability in $\beta$-deformed eigenvalue models}

\author{Aditya Bawane, Pedram Karimi and Piotr Su{\l}kowski}
\emailAdd{abawane@fuw.edu.pl}
\emailAdd{pedram.karimi@fuw.edu.pl}
\emailAdd{psulkows@fuw.edu.pl}

\affiliation{Faculty of Physics, University of Warsaw, ul. Pasteura 5, 02-093 Warsaw, Poland}

\abstract{In this note we provide proofs of various expressions for expectation values of symmetric polynomials in $\beta$-deformed eigenvalue models with quadratic, linear, and logarithmic potentials. The relations we derive are also referred to as superintegrability. Our work completes proofs of superintegrability statements conjectured earlier in literature.
\\
\\
}

\begin{document}

\maketitle
	

\section{Introduction}

In this note we prove the superintegrability property of the $\beta$-deformed eigenvalue models. Superintegrability is the statement that certain quantities in an integrable system can be computed explicitly, in consequence of some extra properties in addition to those that guarantee integrability. A prototype example of such a feature is the existence of closed orbits, described by elementary functions, in the motion in the potential $r^n$ for $n=-1$ and $n=2$, due to an extra conservation law. Recently the superintegrability has been analyzed in much detail in matrix models, where these extra conditions take form of the string equation and Virasoro constraints. One manifestation of superintegrability in this context is the statement that expectation values of various symmetric functions can be explicitly expressed also in terms of analogous symmetric functions evaluated with appropriate arguments \cite{Itoyama:2017xid,Mironov:2017och,Morozov:2018eiq,CLZ,Mironov:2022fsr,Mironov:2022yhd,Mishnyakov:2022bkg,Wang:2022lzj}.  As symmetric functions in this context can be identified with appropriate characters, such relations are often presented schematically as $\langle character \rangle \sim character$. 

Let us summarize our results, and also recent developments concerning superintegrability, in more detail. While several classes of matrix models have been analyzed in this context in literature, in this note we focus on the following $\beta$-deformed eigenvalue model
\begin{equation}
	Z_\beta(V;p_k) \bydef \int \left(\prod_{i=1}^N d z_i\right)\left[\Delta(z)\right]^{2\beta}\exp\left[- \sum_{i=1}^N V(z_i)\right]\exp\left[\beta\sum_{k=1}^{\infty}\frac{p_k}{k} \sum_{i=1}^N z_i^k \right]     \label{mm-intro}
\end{equation}
where $\Delta(z) = \prod_{1\leq i<j\leq N}(z_i-z_j)$ is the Vandermonde factor, and we consider two classes of potentials $V(z)$. The first one is the gaussian potential with a linear term
\begin{equation}
	V(z) = \frac{a_2}{2}z^2 + a_1 z,    \label{gaussian-intro}
\end{equation}
for which we prove the following superintegrability statement, i.e. an explicit expression for the expectation values of Jack polynomials $P_{\lambda}$
\begin{equation}
	\big\langle P_\lambda(z_1,\dots,z_N) \big\rangle = \frac{P_\lambda\{N\}P_\lambda\{\beta^{-1}(a_2\delta_{k,2}-a_1\delta_{k,1})\}}{P_\lambda\{a_2\beta^{-1}\delta_{k,1}\}}. \label{superi-intro}
\end{equation}
We explain details of the notation in the following sections. This statement was conjectured and verified for $|\lambda|\leq9$ in \cite{CLZ}. 
In \cite{Mishnyakov:2022bkg} an outline of a proof was provided in the purely quadratic case $a_1=0$, based on the use of $W$-operators and their actions on a family of symmetric functions pertinent to the model. However some of its crucial steps were only conjectured. In  this note we fill in these gaps and provide a complete proof of (\ref{superi-intro}).\footnote{In the last stage of our work a similar proof, albeit only in a purely quadratic case, was presented in \cite{Wang:2022fxr}.}

Furthermore, we also consider a model with a linear and logarithmic terms in the potential
\begin{equation}
V(z) = -a_1 z + \nu \log z
\end{equation}
and following a similar strategy we prove that
\begin{equation}
	\big\langle P_\lambda(z_1,\dots,z_N) \big\rangle=\frac{P_\lambda\{N\}P_\lambda\{N+\beta^{-1}\nu + \beta^{-1}-1\}}{P_\lambda\{a_1\beta^{-1}\delta_{k,1}\}}.
\end{equation}
This superintegrability statement was conjectured and verified for $|\lambda|\leq9$ in \cite{CLZ}, 
and proven in a purely linear case (i.e. for $\nu=0$) in \cite{Wang:2022lzj}. 

To sum up, in this note we generalize previous superintegrability statements for the $\beta$-deformed eigenvalue models and provide their complete proofs.

\bigskip

Let us also briefly summarize our work from a broader perspective.  In general, our results provide explicit expressions for certain matrix model expectation values. One of the first prominent results of this type is the computation of the Euler characteristics of the moduli space of curves by Harer and Zagier \cite{harer-zagier}, which can be interpreted in terms of enumeration of chord diagrams, or as an explicit computation of the expectation values $\langle \textrm{Tr}\, M^n \rangle$ in the gaussian matrix model that involves integration over matrices $M$. The matrix model interpretation of this calculation was further elucidated by Itzykson and Zuber \cite{itz_zub}. The gaussian model considered in those works corresponds to $\beta=1$ and $p_k=0$ in (\ref{mm-intro}), with the potential (\ref{gaussian-intro}) with $a_1=0$.

More recently, following \cite{Itoyama:2017xid,Mironov:2017och}, expectation values of various symmetric polynomials and the phenomenon referred to as superintegrability started to be analyzed more systematically. 
The results for expectation values of Schur polynomials in the actual hermitian matrix model were subsequently generalized to the $\beta$-deformed eigenvalue ensembles, as we mentioned above \cite{CLZ,Mishnyakov:2022bkg,Wang:2022lzj}. Further deformation that involves two parameters, commonly denoted $q$ and $t$, and corresponding superintegrability statement analogous to (\ref{superi-intro}), however involving Macdonald polynomials, were conjectured in \cite{CLZ,Morozov:2018eiq}. For a review of these developments and some recent results see \cite{Mironov:2022fsr,Mironov:2022yhd}. The methods of the proof that we discuss in this paper do not immediately generalize to the $(q,t)$-deformed case, therefore we leave it for future consideration.  

We also stress that superintegrability is important not only as a property of matrix models, but also in view of their applications in other contexts. On one hand, eigenvalue ensembles and deformations mentioned above arise in gauge theories with extended supersymmetry, in particular in localization of 3-dimensional $\mathcal{N}=2$ theories \cite{CLZ} or in theories with Nekrasov deformation in four dimensions \cite{Dijkgraaf:2009pc,Sulkowski:2009ne}. On the other hand, they arise in the context of knot invariants \cite{Brini:2011wi}. All these relations provide an additional important motivation to derive explicit expressions for expectation values of various symmetric functions.

\bigskip


This paper is organized as follows. In section \ref{sec-gaussian} we provide a proof of superintegrability for the $\beta$-deformed gaussian model with purely quadratic potential. In section \ref{sec-gaussian-linear} we generalize this proof to the case of a gaussian potential with a linear term. In section \ref{sec-linear} we prove superintegrability for a model with a potential with a linear and logarithmic terms.


\section{Gaussian model}   \label{sec-gaussian}

Partition functions of our interest, presented already in (\ref{mm-intro}), involve a potential $V(z)$ and are given by
\begin{equation}\label{Zbeta}
	Z_\beta(V;p_k) \bydef \int \left(\prod_{i=1}^N d z_i\right)\left[\Delta(z)\right]^{2\beta}\exp\left[- \sum_{i=1}^N V(z_i)\right]\exp\left[\beta\sum_{k=1}^{\infty}\frac{p_k}{k} \sum_{i=1}^N z_i^k \right]
\end{equation}
where $\Delta(z) = \prod_{1\leq i<j\leq N}(z_i-z_j)$ is the Vandermonde factor. 
As is a common practice, $p_k$ do double duty of coupling constants in the action, while also denoting power sum polynomials in terms of which various other symmetric functions are expanded. To start with, in this section we consider the purely gaussian potential
\begin{equation}
	V(z) = \frac{a_2}{2}z^2.
\end{equation}

It follows from the invariance of the partition function under the transformation $z_i\rightarrow z_i + \epsilon z_i^{n+1}, n\geq -1, \forall i \in \{1,\dots,N\}$, that the partition function satisfies  the Virasoro constraints:
\begin{equation}
	\mathcal{L}_n Z_\beta = 0, \quad n\geq -1
\end{equation}
where the Virasoro operators $\mathcal{L}_n, n \geq -1$, satisfy $[\mathcal{L}_n,\mathcal{L}_m] = (n-m)\mathcal{L}_{n+m}$. Explicitly,
\begin{equation}
	\mathcal{L}_n = -\frac{a_2(n+2)}{\beta}\frac{\partial}{\partial p_{n+2}} + L_n
\end{equation}
where $L_n$'s are the potential-independent part of the Virasoro operators:
\begin{equation}\label{Virasoro}
	\begin{aligned}
		L_{-1}&= \beta p_1 N + \sum_{k=1}^{\infty}kp_{k+1}\frac{\partial}{\partial p_k},\\
		L_{0}&=  \beta N^2 + (1-\beta)N + \sum_{k=1}^{\infty}kp_{k}\frac{\partial}{\partial p_k},\\
		L_{n> 0}&= \sum_{k=1}^{\infty}(n+k)p_{k}\frac{\partial}{\partial p_{n+k}} + 2Nn\frac{\partial}{\partial p_{n}}+ \\&\quad +\frac{1}{\beta}\sum_{k=1}^{n-1}k(n-k)\frac{\partial^2}{\partial p_{k}\partial p_{n-k}} + \frac{1-\beta}{\beta}n(n+1)\frac{\partial}{\partial p_{n}}.
	\end{aligned}
\end{equation}
It follows \textit{a fortiori} that
\begin{equation}\label{singleVir}
	-a_2^{-1}\beta \sum_{n=1}^{\infty}p_n\mathcal{L}_{n-2}Z_\beta \bydef \left(\mathcal{D} - a_2^{-1}W_{-2}\right)Z_\beta = 0,
\end{equation}
where we have defined $\mathcal{D} \bydef \sum_{n=1}^{\infty}n p_n\frac{\partial}{\partial p_{n}}$ and
\begin{equation}\label{Wm2}
	\begin{aligned}
	W_{-2}\bydef &\beta^2 N p_1^2  + \beta^2 N^2 p_2 + (1-\beta)\beta N p_2 +\\
	&+ \sum_{k,l=1}^{\infty}\left[\beta(k+l-2)p_kp_l\frac{\partial}{\partial p_{k+l-2}} + klp_{k+l+2}\frac{\partial^2}{\partial p_{k}\partial p_{l}}\right] + \\
	&+\sum_{k=1}^{\infty}\left[2\beta N k p_{k+2} \frac{\partial}{\partial p_{k}} + (1-\beta)k(k+1)p_{k+2}\frac{\partial}{\partial p_{k}}\right].
\end{aligned}\end{equation}
It can be verified that $[\mathcal{D}, W_{-2}] = 2 W_{-2}$. The partition function \eqref{Zbeta} can be then written as
\begin{equation}\label{ZW}
	Z_\beta = e^{W_{-2}/2a_2}\cdot 1.
\end{equation}

Our goal is to evaluate the right-hand side above. It turns out that the Jack polynomials constitute a natural basis of special functions to describe the action of $W_{-2}$.

We denote Jack polynomials in the so-called P normalization as $P_\lambda$. This notation is standard in literature on symmetric polynomials, see e.g. \cite{macdonald1998symmetric}. These are precisely the ones denoted as $J_\lambda$ in recent works on superintegrability \cite{Mishnyakov:2022bkg,Wang:2022lzj}. However, the reader should note that in the literature on symmetric functions, $J_\lambda$ is used for Jack polynomials in the so-called J normalization, also known as the integral form of Jack polynomials. The latter ones are rarely used in this work.

When a symmetric function, $P_\lambda$ for instance, is written in terms of the basic variables, we will denote its arguments within parentheses: $P_\lambda(z_1,\dots,z_N)$. More often, we will need to write them in terms of the power sum polynomials $p_k= \sum_{i=1}^{N}z_i^k$, in which case we will denote them as $P_\lambda\{p_k\}$. In particular, $P_\lambda\{f(k)\}$ is a short-hand for $P_\lambda\{p_k=f(k)\}$.

Jack polynomials are known to be eigenfunctions of the following operators \cite{Cai:2011jv}:
\begin{equation}\label{DH}
	\begin{aligned}
		\mathcal{D} &= \sum_{k=1}^{\infty}kp_{k}\frac{\partial}{\partial p_{k}},\\
		\mathcal{H} &= \frac{1}{2}\sum_{k,l=1}^{\infty}\left[\beta(k+l)p_kp_l\frac{\partial}{\partial p_{k+l}} + klp_{k+l}\frac{\partial^2}{\partial p_{k}\partial p_{l}}\right]
		+\frac{1-\beta}{2}\sum_{k=1}^{\infty}k(k-1)p_{k}\frac{\partial}{\partial p_{k}},
	\end{aligned}
\end{equation}
so that
\begin{equation}
	\begin{aligned}
		\mathcal{D}P_\lambda &= |\lambda|P_\lambda,\\
		\mathcal{H}P_\lambda &= c^{(\beta)}_\lambda P_\lambda,
	\end{aligned}
\end{equation}
where $|\lambda| = \sum_i\lambda_i$ is the number of boxes in the partition $\lambda$ and
\begin{equation}
	c^{(\beta)}_\lambda = \sum_{(i,j)\in \lambda}(j-1-\beta(i-1)) = n(\lambda')- \beta n(\lambda).
\end{equation}
To our knowledge, the earliest version of this result appears as Theorem 3.1 in \cite{STANLEY198976} (whose author, R. P. Stanley, attributes it to I. G. Macdonald), but the operator is presented in terms of the basic variables rather than the power sum polynomials.

Also, note that the operator $\mathcal{H}$ defined above differs from a similar operator defined in \cite{Wang:2022lzj} 
and \cite{Cai:2011jv} 
by a shift of $\mathcal{D}$. This makes the eigenvalues of the resulting operator more natural, while also simplifying the expression \eqref{wm2_comm} for $W_{-2}$ given below.

A crucial step in the proof is determining the action of $W_{-2}$ on Jack polynomials. In \cite{Wang:2022lzj}, $W_{-2}$ was presented in terms of $\mathcal{H}$ when $\beta =1$, which lets us almost immediately read off the action of $W_{-2}$ on the Schur polynomials. Following a similar strategy for an indeterminate $\beta$, one finds
\begin{equation}\label{wm2_comm}
	\begin{aligned}
		W_{-2} = &\frac{1}{4}\left[\mathcal{H},\left[\mathcal{H},p_2\right]\right]+N\beta \left[\mathcal{H},p_2\right] + \\& + N^2\beta^2p_2-\frac{1}{4}(1-\beta+\beta^2)p_2-\frac{\beta(1-\beta)}{4}p_1^2 .
	\end{aligned}
\end{equation}
We will also need the following Pieri rule for Jack polynomials (see Theorem 6.3 of \cite{knopsahi}, and \cite{naqvi} for additional explanations and examples):
\begin{equation}\label{pieri_column}
	P_{(1^r)}P_\mu = \sum_\lambda c^\lambda_{\mu,(1^r)} P_\lambda
\end{equation}
where the coefficients $c^\lambda_{\mu,(1^r)}$ are known and to which we will return later. For now, we simply note that the sum on the right runs over partitions $\lambda$ such that $\lambda-\mu$ is a vertical $r$-strip (i.e. a skew diagram consisting of $r$ boxes, no two of which are in the same row).  We also note that $P_{(1^r)} = s_{(1^r)} = e_r$, where $s$ and $e$ are the Schur and elementary symmetric polynomials respectively.

There is a similar Pieri rule for Jack polynomials indexed by row partitions  (see Proposition 5.3 and Theorem 6.1 in \cite{STANLEY198976}), given in terms of the $J_\lambda$:
\begin{equation}\label{pieri_row}
	J_{r}J_\mu = \sum_\lambda \frac{g^\lambda_{\mu,r}}{j_\lambda} J_\lambda
\end{equation}
where now the sum runs over partitions $\lambda$ such that $\lambda-\mu$ is a horizontal $r$-strip (i.e. a skew diagram consisting of $r$ boxes, no two of which are in the same column). The constants $g^\lambda_{\mu,r}$ and $j_\lambda$ are known, but we will not need to use them.

In particular, for $r=1$, the diagrams $\lambda$ on the right are those obtained by adding one box to $\mu$. Multiplying two factors of $P_1 = p_1$ with $P_\mu$ therefore gives a sum over all partitions where two boxes are added to $\mu$ without any constraint. For $r=2$, multiplying with $P_{(1,1)} = (p_1^2 - p_2)/2$ gives a sum over partitions, where the two additional squares cannot be in the same row. Consequently, expanding $p_2 P_\mu$ over Jack polynomials gives a sum over partitions where two boxes are added to $\mu$ without any constraint. One may therefore write
\begin{equation}\label{p2P}
	p_2 P_\mu = \sum_{\lambda = \mu +\square + \square} C_{\mu\lambda} P_{\lambda}.
\end{equation}
One may also write
\begin{equation}
	p_1^2 P_\mu = \sum_{\lambda = \mu +\square + \square} A_{\mu\lambda} C_{\mu\lambda} P_{\lambda}.
\end{equation}
Using this notation and \eqref{wm2_comm}, the action of $W_{-2}$ on a Jack polynomial is:
\begin{equation}
	\begin{aligned}
		W_{-2}P_\mu = \sum_{\lambda = \mu +\square_1 + \square_2} \bigg[&(j_1-1+\beta(N-i_1+1))(j_2-1+\beta(N-i_2+1)) +\\ &+\frac{1}{4}(j_2-j_1+\beta(i_1-i_2))^2 - \frac{1}{4}(1-\beta+\beta^2)+\\&-\frac{\beta(1-\beta)}{4}A_{\mu\lambda}\bigg]C_{\mu\lambda} P_{\lambda},
	\end{aligned}
\end{equation}
where $(i_1,j_1)$ and $(i_2,j_2)$ are coordinates of boxes $\square_1$ and $\square_2$ respectively.
We observe that the first line on the right-hand side corresponds precisely to the expression in eqn. (35) of \cite{Mishnyakov:2022bkg} (noting that they count rows and columns starting from 0, while we begin counting from 1). We now show, by computing $A_{\mu\lambda}$, that the terms in the second and third rows cancel.

Observe that
\begin{equation}
	A_{\mu\lambda} = \frac{\langle P_{\lambda}, p_1^2 P_\mu\rangle}{\langle P_{\lambda}, p_2 P_\mu\rangle},
\end{equation}
where $\langle\cdot,\cdot\rangle$ denotes the Jack scalar product. The constraint in the summation in \eqref{pieri_column} tells us that $\langle P_{\mu + \twoh}, P_{(1,1)} P_\mu\rangle =0$,  implying that $A_{\mu,\mu + \twoh}=1$. Likewise, \eqref{pieri_row} tells us that $\langle J_{\mu + \twov}, J_{2} J_\mu\rangle =0$. Using $J_2 =  p_1^2 + \beta^{-1}p_2$ we see that
\begin{equation}
	-\beta = \frac{\langle J_{\mu + \twov}, p_1^2 J_\mu\rangle}{\langle J_{\mu + \twov}, p_2 J_\mu\rangle} = \frac{\langle P_{\mu + \twov}, p_1^2 P_\mu\rangle}{\langle P_{\mu + \twov}, p_2 P_\mu\rangle} = A_{\mu,\mu + \twov}.
\end{equation}
More generally, one can write
\begin{equation}
	A_{\mu\lambda} = \frac{\langle P_{\lambda}, p_1^2 P_\mu\rangle}{\langle P_{\lambda}, p_2 P_\mu\rangle} = \frac{1}{1-2\frac{\langle P_{\lambda}, P_{(1,1)} P_\mu\rangle}{\langle P_{\lambda}, P_1^2 P_\mu\rangle}}.	
\end{equation}
We consider now the case where the two boxes are staggered, i.e. added to neither the same row nor the same column. The above expression can be related to the coefficients appearing in the Pieri rule \eqref{pieri_column} as
\begin{equation}\label{jack_prod_ratio}
	\frac{\langle P_{\lambda}, P_1^2 P_\mu\rangle}{\langle P_{\lambda}, P_{(1,1)} P_\mu\rangle} = \frac{\sum_{\sigma=\mu+\square}c^\lambda_{\sigma,1}c^\sigma_{\mu,1}}{c^\lambda_{\mu,(1,1)} } = \frac{c^{\mu+\square_1+\square_2}_{\mu+\square_2,1}c^{\mu+\square_2}_{\mu,1}+ c^{\mu+\square_1+\square_2}_{\mu+\square_1,1}c^{\mu+\square_1}_{\mu,1}}{c^{\mu+\square_1+\square_2}_{\mu,(1,1)}}.
\end{equation}
The sum over $\sigma$ in the middle expression consists of two terms, corresponding to the order in which the two boxes are added to reach the same final shape $\lambda =\mu+\square_1+\square_2$. Without loss of generality, we may label the box to the left as box 1 $(\square_1 = (i_1,j_1))$ and the one to the right as box 2 $(\square_2= (i_2,j_2))$, so that $i_1 > i_2$ and $j_1<j_2$.

We now need the explicit form of the coefficients:
\begin{equation}\label{piericoeff}
	c^\lambda_{\mu,(1^r)} = \prod_{s\in X(\lambda/\mu)} \frac{h_*^\lambda(s)h^*_\mu(s)}{h_*^\mu(s)h^*_\lambda(s)}
\end{equation}
where $X(\lambda/\mu)$ is the set of boxes $(i,j) \in \mu$ such that $\mu_i = \lambda_i$ and $\mu_j < \lambda_j$. For example, in the figure below, $\mu = (7,5,3,3,1,1)$, and the yellow squares are the ones that have been added to obtain various $\lambda$'s, while the blue squares constitute the set  $X(\lambda/\mu)$. Note how on adding two boxes together, as in the last figure, we get one blue box less than the union of the set of blue boxes when only one box is added at a time.
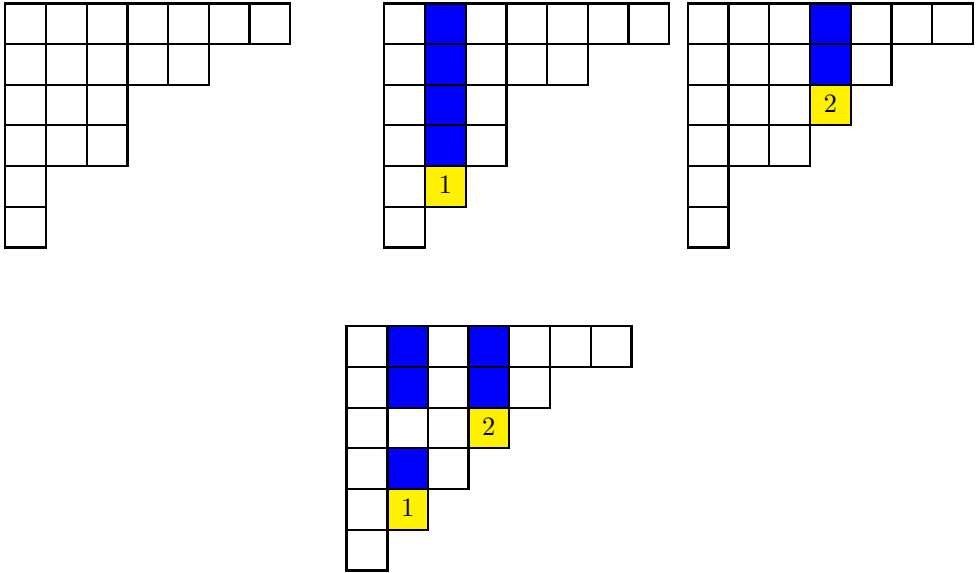
\begin{center}
	\ytableausetup{centertableaux}
	\begin{ytableau}
		{} & {} & {} & {} & {} & {} & {} \\
		{} & {} & {} & {} & {}  \\
		{} & {} & {}  \\
		{} & {} & {}  \\
		{}   \\
		{}  \\
	\end{ytableau}\hspace{1cm}
	\ytableausetup{centertableaux}
	\begin{ytableau}
	{} & *(blue) & {} & {} & {} & {} & {} \\
	{} & *(blue) & {} & {} & {}  \\
	{} & *(blue) & {}  \\
	{} &  *(blue)& {}  \\
	{} & *(yellow)1 \\
	{}  \\
\end{ytableau}\vspace{1cm}
	\ytableausetup{centertableaux}
\begin{ytableau}
	{} &  & {} & *(blue) & {} & {} & {} \\
	{} &  & {} & *(blue) & {}  \\
	{} &  & {}  & *(yellow) 2\\
	{} & & {}  \\
	{}  \\
	{}  \\
\end{ytableau}\hspace{1cm}
	\ytableausetup{centertableaux}
\begin{ytableau}
	{} & *(blue) & {} & *(blue) & {} & {} & {} \\
	{} & *(blue) & {} & *(blue) & {}  \\
	{} &  & {}  & *(yellow) 2\\
	{} &  *(blue)& {}  \\
	{} & *(yellow)1 \\
	{}  \\
\end{ytableau}
\end{center}
Furthermore, in (\ref{piericoeff}), $h^*_\lambda(s)$ and $h_*^\lambda(s)$ are respectively the upper and lower hook lengths of the box $s$:
\begin{equation}
	\begin{aligned}
		h^*_\lambda(s) &= \beta^{-1}(a_\lambda(s)+1)+l_\lambda(s)\\
		h_*^\lambda(s) &= \beta^{-1}a_\lambda(s)+l_\lambda(s)+1
	\end{aligned}
\end{equation} 
where $a_\lambda(i,j) = \lambda_i - j$ is the arm length and  $l_\lambda(i,j) = \lambda_j' - i$ is the leg length of $(i,j)\in \lambda$. (A useful mnemonic is that in the definition of the lower hook length of a given box, the box itself is treated as part of the leg, going downwards.) For example, the yellow square in the diagram below has coordinates $(i,j)= (2,3)$, with $a_\lambda = 4$, $l_\lambda = 2$, $h^*_\lambda = 5\beta^{-1} + 2$, and $h_*^\lambda = 4\beta^{-1} + 3$.
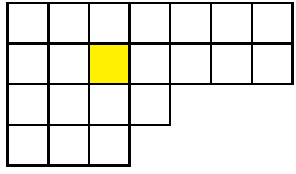
\begin{center}
	\ytableausetup{centertableaux}
	\begin{ytableau}
		{} & {} & {} & {} & {} & {} & {} \\
		{} & {} & *(yellow) & {} & {} & {}& {}  \\
		{} & {} & {}& {}  \\
		{} & {} & {}
	\end{ytableau}
\end{center}

Consider now the first term of \eqref{jack_prod_ratio}, where the box on the right $(\square_2)$ is added first:
\begin{equation}
	t_1(\square_1,\square_2) \bydef \frac{c^{\mu+\square_1+\square_2}_{\mu+\square_2,1}c^{\mu+\square_2}_{\mu,1}}{c^{\mu+\square_1+\square_2}_{\mu,(1,1)}}.
\end{equation}
Due to the product form of \eqref{piericoeff}, one can expect a lot of cancellations in the above expression. However, exactly one factor survives: the contribution to $c^{\mu+\square_1+\square_2}_{\mu+\square_2,1}$ due to the box $(i_2,j_1)$, which does not appear in the denominator. Therefore
\begin{equation}
	\begin{aligned}
		t_1(\square_1,\square_2) &= \frac{h_*^{\mu+\square_1+\square_2}(i_2,j_1)h^*_{\mu+\square_2}(i_2,j_1)}{h_*^{\mu+\square_2}(i_2,j_1)h^*_{\mu+\square_1+\square_2}(i_2,j_1)}\\
		&=\frac{\left[\beta^{-1}(j_2-j_1)+i_1-i_2+1\right]\left[\beta^{-1}(j_2-j_1+1)+i_1-i_2-1\right]}{\left[\beta^{-1}(j_2-j_1+1)+i_1-i_2\right]\left[\beta^{-1}(j_2-j_1)+i_1-i_2\right]}.
\end{aligned}\end{equation}
(It helps to note that $i_1 = \mu'_{j_1}+1$, $j_1 = \mu_{i_1}+1$, $i_2 = \mu'_{j_2}+1$, and $j_2 = \mu_{i_2}+1$.)
In the second term of \eqref{jack_prod_ratio}, where the left box $(\square_1)$ is added first
\begin{equation}
	t_2(\square_1,\square_2) \bydef \frac{c^{\mu+\square_1+\square_2}_{\mu+\square_1,1}c^{\mu+\square_1}_{\mu,1}}{c^{\mu+\square_1+\square_2}_{\mu,(1,1)}}
\end{equation}
one has similar cancellations, leaving a contribution to $c^{\mu+\square_1}_{\mu,1}$ due to the box $(i_2,j_1)$:
\begin{equation}
	\begin{aligned}
		t_2(\square_1,\square_2) &= \frac{h_*^{\mu+\square_1}(i_2,j_1)h^*_{\mu}(i_2,j_1)}{h_*^{\mu}(i_2,j_1)h^*_{\mu+\square_1}(i_2,j_1)}\\
		&=\frac{\left[\beta^{-1}(j_2-j_1-1)+i_1-i_2+1\right]\left[\beta^{-1}(j_2-j_1)+i_1-i_2-1\right]}{\left[\beta^{-1}(j_2-j_1)+i_1-i_2\right]\left[\beta^{-1}(j_2-j_1-1)+i_1-i_2\right]}.
\end{aligned}\end{equation}
We digress momentarily to point out that $t_1(\square_1,\square_2) = t_2(\square_2,\square_1)$. This is not \textit{a priori} obvious because the labels 1 and 2 were assigned based on the relative positions of the boxes, and indeed the intermediate expressions in terms of the hook lengths $h^*$ and $h_*$ are not manifestly equal under this exchange.

We now have the expression for $A_{\mu\lambda}$:
\begin{equation}\label{key}
	A_{\mu\lambda} = \frac{\beta-1+(j_2-j_1+\beta(i_1-i_2+1))(j_2-j_1+\beta(i_1-i_2-1))}{\beta(1-\beta)}.
\end{equation}
While we derived this with the assumption that $\lambda-\mu$ was neither a row nor a column, we see that these two cases are also captured by this expression, as can be seen by setting $i_1=i_2, j_2-j_1=1$ or $j_1=j_2, i_1-i_2=1$ for a row or a column respectively. This is therefore the general expression.

It is  now straightforward to see that
\begin{equation}
	\beta(1-\beta)A_{\mu\lambda} = (j_2-j_1+\beta(i_1-i_2))^2 - (1-\beta+\beta^2),
\end{equation}
thus proving 
\begin{equation}\label{Waction}
		W_{-2}P_\mu = \sum_{\lambda = \mu +\square_1 + \square_2} (j_1-1+\beta(N-i_1+1))(j_2-1+\beta(N-i_2+1))C_{\mu\lambda} P_{\lambda}.
\end{equation}

To calculate \eqref{ZW}, we need to act with $W_{-2}$ multiple times on $P_{\phi}=1$, which can now be done using \eqref{Waction}:
\begin{equation}\label{Wn1}
	\begin{aligned}
	(W_{-2})^n\cdot 1 = \sum_{\lambda = \lambda^{(2n)}}\Big[&\prod_{(i,j)\in \lambda^{(2n)}}(j-1+\beta(N-i+1))\\ 
	&\times\sum_{\lambda^{(2)}\dots\lambda^{(2n-2)}} C_{\phi\lambda^{(2)}}C_{\lambda^{(2)}\lambda^{(4)}}\dots C_{\lambda^{(2n-2)}\lambda^{(2n)}}P_{\lambda}\Big]
\end{aligned}\end{equation}
where $\lambda^{(2m)}$ is obtained by adding 2 boxes successively $m$-times starting from the initial empty partition $\phi$, so that  $|\lambda^{(2m)}| = 2m$. The rest of the proof consists of writing the product over the cells in the first line, and the sum of products of $C$'s in the second line, in terms of Jack polynomials $P_\lambda\{p_k\}$ evaluated at special values of the power sum polynomials $p_k$.

We begin with the $C$'s. The Cauchy identity for Jack polynomials reads
\begin{equation}\label{cauchy}
	\exp\left[\beta \sum_{k=1}^{\infty} \frac{p_k\bar{p}_k}{k}\right] = \sum_{\lambda}\frac{P_\lambda\{p_k\}P_\lambda\{\bar{p}_k\}}{\langle P_\lambda,P_\lambda\rangle}.
\end{equation}
Setting $\bar{p}_k = a_2\beta^{-1}\delta_{k,2}$ we see that
\begin{equation}
	\exp\left[{\frac{a_2p_2}{2}}\right] = \sum_{\lambda}\frac{P_\lambda\cdot P_\lambda\{a_2\beta^{-1}\delta_{k,2}\}}{\langle P_\lambda,P_\lambda\rangle}.
\end{equation}
On the other hand, directly by the definition of $C_{\mu\lambda}$ in \eqref{p2P}, we see that
\begin{equation}
	\exp\left[\frac{a_2p_2}{2}\right] = \sum_{n\geq 0}\frac{a_2^{n} P_{\lambda^{(2n)}}}{2^{n}n!}\sum_{\lambda^{(2)}\dots\lambda^{(2n-2)}} C_{\phi\lambda^{(2)}}C_{\lambda^{(2)}\lambda^{(4)}}\dots C_{\lambda^{(2n-2)}\lambda^{(2n)}}.
\end{equation}
The above expressions together give
\begin{equation}\label{prodC}
	\sum_{\lambda^{(2)}\dots\lambda^{(2n-2)}; \lambda^{(2n)}=\lambda} C_{\phi\lambda^{(2)}}C_{\lambda^{(2)}\lambda^{(4)}}\dots C_{\lambda^{(2n-2)}\lambda^{(2n)}} = \frac{2^{|\lambda|/2}(|\lambda|/2)!}{a_2^{|\lambda|/2}}\frac{P_\lambda\{a_2\beta^{-1}\delta_{k,2}\}}{\langle P_\lambda,P_\lambda\rangle}.
\end{equation}
For the first line of \eqref{Wn1}, we turn to \cite{macdonald1998symmetric}. By eqn. (10.20) in this reference
\begin{equation}
	P_\lambda\{N\} = \prod_{(i,j)\in\lambda}\frac{\beta^{-1}(j-1)+ N - i + 1}{h_*^\lambda(i,j)},
\end{equation}
while eqns. (10.22) and (10.29) tell us that
\begin{equation}
	P_\lambda\{\delta_{k,1}\} = \prod_{(i,j)\in\lambda}\frac{1}{h_*^\lambda(i,j)}.
\end{equation}
Putting these together, we can write
\begin{equation}\label{prodcells}
	 \prod_{(i,j)\in \lambda}(j-1+\beta(N-i+1)) = \frac{P_\lambda\{N\}}{P_\lambda\{\beta^{-1}\delta_{k,1}\}}.
\end{equation}
We now have all the ingredients to write \eqref{ZW} entirely in terms of Jack polynomials:
\begin{equation}
	Z_\beta(p_k) = e^{W_{-2}/2a_2}\cdot 1 = \sum_\lambda \frac{P_\lambda\{p_k\}}{\langle P_\lambda,P_\lambda\rangle}\frac{P_\lambda\{N\}P_\lambda\{a_2\beta^{-1}\delta_{k,2}\}}{P_\lambda\{a_2\beta^{-1}\delta_{k,1}\}}.
\end{equation}
(Note that $P_\lambda\{x\delta_{k,2}\} = x^{|\lambda|/2}P_\lambda\{\delta_{k,2}\}$ and $P_\lambda\{x\delta_{k,1}\} = x^{|\lambda|}P_\lambda\{\delta_{k,1}\} $.) On the other hand, again by the Cauchy identity \eqref{cauchy}, one can expand the partition function \eqref{Zbeta} as
\begin{equation}\label{Zcorr}
	Z_\beta(V;p_k) = \sum_\lambda \frac{P_\lambda\{p_k\}}{\langle P_\lambda,P_\lambda\rangle}\langle P_\lambda(z_1,\dots,z_N) \rangle,
\end{equation}
regardless of the potential in question.
The orthogonality of the Jack polynomials gives us the desired result
\begin{equation}
	\big\langle P_\lambda(z_1,\dots,z_N) \big\rangle = \frac{P_\lambda\{N\}P_\lambda\{a_2\beta^{-1}\delta_{k,2}\}}{P_\lambda\{a_2\beta^{-1}\delta_{k,1}\}}.
\end{equation}


\section{Gaussian model with a linear term}    \label{sec-gaussian-linear}

We now consider a slightly more general potential
\begin{equation}
	V(z) = \frac{a_2}{2}z^2 + a_1 z
\end{equation}
and denote the corresponding partition function (\ref{Zbeta}) as $Z_\beta(a_2, a_1; p_k)$. The Virasoro operators annihilating the partition function are modified to
\begin{equation}
	\mathcal{L}_n = -\frac{a_2(n+2)}{\beta}\frac{\partial}{\partial p_{n+2}} -\frac{a_1(n+1)}{\beta}\frac{\partial}{\partial p_{n+1}}-a_1 N \delta_{n,-1} + L_n, \quad n\geq -1,
\end{equation}
where $L_n$'s are given in \eqref{Virasoro}. The Virasoro constraints now imply the single equation\begin{equation}
	\left(\mathcal{D} - \frac{W_{-2}}{a_2} + \frac{a_1L_{-1}}{a_2}\right)Z_\beta(a_2, a_1; p_k) = 0,
\end{equation}
where $W_{-2}$ is the same as in \eqref{Wm2}, and $[\mathcal{D},L_{-1}]=L_{-1}$. A crucial observation is that $[W_{-2},L_{-1}]=0$. The partition function can now be written as
\begin{equation}\label{Zexp}
	Z_\beta(a_2, a_1; p_k) = \exp\left[\frac{W_{-2}}{2a_2} - \frac{a_1L_{-1}}{a_2}\right]\cdot 1 = e^{W_{-2}/2a_2}\cdot e^{-a_1L_{-1}/a_2}\cdot 1.
\end{equation}
We now follow a similar strategy as in the previous section: we work out the action of $L_{-1}$ on Jack polynomials. This, as we have seen, is facilitated by writing it in terms of the operator $\mathcal{H}$, defined in \eqref{DH}, which has the Jack polynomials as its eigenfunctions:
\begin{equation}\label{Lm1comm}
	L_{-1} = [\mathcal{H},p_1]+\beta N p_1.
\end{equation}
Analogously to \eqref{p2P}, let us define constants $B_{\mu\lambda}$ as coefficients in the expansion
\begin{equation}\label{p1P}
	p_1 P_\mu = \sum_{\lambda = \mu +\square} B_{\mu\lambda} P_{\lambda}.
\end{equation}
These constants are of course known due to \eqref{piericoeff}, but we will not need their explicit expression, just as in the case of $C_{\mu\lambda}$ in the previous section. Using \eqref{Lm1comm} and \eqref{p1P} we get:
\begin{equation}\label{Laction}
	L_{-1}P_\mu = \sum_{\lambda = \mu +\square} (j-1+\beta(N-i+1))B_{\mu\lambda} P_{\lambda},
\end{equation}
where $(i,j)$ are coordinates of the additional box $\square$.

On expanding the exponentials in the right-most side of \eqref{Zexp}, one has to deal with terms of the type
\begin{equation}\label{WnLm}
	\begin{aligned}
		(W_{-2})^n(L_{-1})^m\cdot 1 &= \sum_{\lambda = \lambda^{(2n+m)}}\Big[\prod_{(i,j)\in \lambda^{(2n+m)}}(j-1+\beta(N-i+1))\\ 
		&\quad\times\sum_{\lambda^{(1)}\dots\lambda^{(m)}} B_{\phi\lambda^{(1)}}B_{\lambda^{(1)}\lambda^{(2)}}\dots B_{\lambda^{(m-1)}\lambda^{(m)}}\\
		&\quad\times\sum_{\lambda^{(m+2)}\dots\lambda^{(m+2n-2)}} C_{\lambda^{(m)}\lambda^{(m+2)}}\dots C_{\lambda^{(m+2n-2)}\lambda^{(m+2n)}} P_{\lambda^{(m+2n)}}\Big].
\end{aligned}\end{equation}
Setting $\bar{p}_k = a_2\beta^{-1}\delta_{k,2} - a_1\beta^{-1}\delta_{k,1}$ in the Cauchy identity \eqref{cauchy}, we see that
\begin{equation}
	\exp\left[{\frac{a_2p_2}{2}}-a_1p_1\right] = \sum_{\lambda}\frac{P_\lambda\cdot P_\lambda\{\beta^{-1}(a_2\delta_{k,2}-a_1\delta_{k,1})\}}{\langle P_\lambda,P_\lambda\rangle}.
\end{equation}
On the other hand, by definition of $B_{\mu\lambda}$ and $C_{\mu\lambda}$
\begin{equation}
	\begin{aligned}
	&\exp\left[{\frac{a_2p_2}{2}}-a_1p_1\right] = \\ &\qquad=\sum_{n,m\geq 0} \Big[P_{\lambda^{(m+2n)}} \frac{\left(a_2/2\right)^n(-a_1)^m}{n!m!}\sum_{\lambda^{(1)}\dots\lambda^{(m)}} B_{\phi\lambda^{(1)}}B_{\lambda^{(1)}\lambda^{(2)}}\dots B_{\lambda^{(m-1)}\lambda^{(m)}}\\ &\qquad\qquad\times\sum_{\lambda^{(m+2)}\dots\lambda^{(m+2n-2)}}  C_{\lambda^{(m)}\lambda^{(m+2)}}\dots C_{\lambda^{(m+2n-2)}\lambda^{(m+2n)}}\Big].
\end{aligned}\end{equation}
The above two equations together imply, for $\lambda = \lambda^{(m+2n)}$,
\begin{equation}
	\begin{aligned}
		&\frac{ P_\lambda\{\beta^{-1}(a_2\delta_{k,2}-a_1\delta_{k,1})\}}{\langle P_\lambda,P_\lambda\rangle} =\\
		&\qquad=\frac{\left(a_2/2\right)^n(-a_1)^m}{n!m!}\sum_{\lambda^{(1)}\dots\lambda^{(m)}} B_{\phi\lambda^{(1)}}B_{\lambda^{(1)}\lambda^{(2)}}\dots B_{\lambda^{(m-1)}\lambda^{(m)}}\\ &\qquad \qquad\times\sum_{\lambda^{(m+2)}\dots\lambda^{(m+2n-2)}}  C_{\lambda^{(m)}\lambda^{(m+2)}}\dots C_{\lambda^{(m+2n-2)}\lambda^{(m+2n)}}
	\end{aligned}
\end{equation}
Summing \eqref{WnLm} over $n$ and $m$ with appropriate coefficients, while also recalling \eqref{prodcells}, we can finally write
\begin{equation}
	\begin{aligned}
		Z_\beta(a_2, a_1; p_k) &= e^{W_{-2}/2a_2}\cdot e^{-a_1L_{-1}/a_2}\cdot 1\\&= \sum_\lambda \frac{P_\lambda\{p_k\}}{\langle P_\lambda,P_\lambda\rangle}\frac{P_\lambda\{N\}P_\lambda\{\beta^{-1}(a_2\delta_{k,2}-a_1\delta_{k,1})\}}{P_\lambda\{a_2\beta^{-1}\delta_{k,1}\}}.
	\end{aligned}
\end{equation}
Using \eqref{Zcorr}, we get the desired result for the correlators
\begin{equation}
	\big\langle P_\lambda(z_1,\dots,z_N) \big\rangle = \frac{P_\lambda\{N\}P_\lambda\{\beta^{-1}(a_2\delta_{k,2}-a_1\delta_{k,1})\}}{P_\lambda\{a_2\beta^{-1}\delta_{k,1}\}}.
\end{equation}
This relation was conjectured and verified for $|\lambda|\leq9$ in \cite{CLZ}, and here we provided its complete proof. 


\section{Linear and logarithmic potential}   \label{sec-linear}

Finally, following an analogous strategy, we consider the eigenvalue model (\ref{Zbeta}) with a potential with a linear and logarithmic terms. Superintegrability property of a linear theory $V(z) = z$ of $N\times N$ complex matrices was proven in \cite{Wang:2022lzj}. We generalize that proof and include a determinant factor in the action, which can  be also interpreted as an extra logarithmic term in the potential:
\begin{equation}\label{Zdet}
		Z_\beta \bydef \int_0^{\infty} \left(\prod_{i=1}^N d z_i\right)\left[\Delta(z)\right]^{2\beta} \left(\prod_{i=1}^N  z_i^\nu\right) e^{- a_1 \sum_{i=1}^N z_i}\exp\left[\beta\sum_{k=1}^{\infty}\frac{p_k}{k} \sum_{i=1}^N z_i^k \right].
\end{equation}
This integral can be viewed as the $\beta$-deformation of the (anti-)Wishart model over the space of complex matrices of size $N_1\times N_2$, with $N=\min(N_1,N_2)$ and $\nu = |N_1-N_2|$, see e.g. chapter 13 of \cite{livan}.

The Virasoro operators that annihilate the above partition function are given by 
\begin{equation}
	\mathcal{L}_n = -\frac{a_1(n+1)}{\beta}\frac{\partial}{\partial p_{n+1}}-a_1 N \delta_{n,-1}+\nu\left(\frac{n}{\beta}\frac{\partial}{\partial p_n} + N\delta_{n,0}\right) + L_n, \quad n\geq -1.
\end{equation}
The single constraint in this case is $(\mathcal{D} -a_1^{-1} W_{-1})Z_\beta = 0$, so that $Z_\beta = e^{W_{-1}/a_1}\cdot1$,  where
\begin{equation}\label{Wm1}
	\begin{aligned}
		W_{-1}\bydef &\beta N (\nu + \beta N  +1 -\beta) p_1 + \\
		&+ \sum_{k,l=1}^{\infty}\left[\beta(k+l-1)p_kp_l\frac{\partial}{\partial p_{k+l-1}} + klp_{k+l+1}\frac{\partial^2}{\partial p_{k}\partial p_{l}}\right] + \\
		&+\sum_{k=1}^{\infty}(k-1)(\nu + 2\beta N  +k -k\beta)p_{k}\frac{\partial}{\partial p_{k-1}},
\end{aligned}\end{equation}
which can be written as
\begin{equation}\label{wm1_comm}
	\begin{aligned}
		W_{-1} = &\left[\mathcal{H},\left[\mathcal{H},p_1\right]\right]+(1-\beta + \nu + \beta N) \left[\mathcal{H},p_1\right] + \\& + \beta N (\nu + \beta N  +1 -\beta) p_1.
	\end{aligned}
\end{equation}
This lets us write
\begin{equation}\label{Wm1action}
	\begin{aligned}
	W_{-1}P_\mu = \sum_{\lambda = \mu +\square} \Big[&\left(j-1+\beta(N-i+1)\right)\\ &\times\left(j-1+\beta((N+\beta^{-1}\nu + \beta^{-1}-1)-i+1)\right)B_{\mu\lambda} P_{\lambda}\Big],
\end{aligned}\end{equation}
where $(i,j)$ are coordinates of $\square$. Once again, expanding $e^{W_{-1}/a_1}\cdot1$, we find terms of the type
\begin{equation}\label{Wm1n}
	\begin{aligned}
	W_{-1}^n\cdot 1 &= \sum_{\lambda = \lambda^{(n)}}\Big[\prod_{(i,j)\in \lambda^{(n)}}(j-1+\beta(N-i+1))\\ 
	&\quad\times\prod_{(i,j)\in \lambda^{(n)}}\left(j-1+\beta((N+\beta^{-1}\nu + \beta^{-1}-1)-i+1)\right)\\ 
	&\quad\times\sum_{\lambda^{(1)}\dots\lambda^{(n-1)}} B_{\phi\lambda^{(1)}}B_{\lambda^{(1)}\lambda^{(2)}}\dots B_{\lambda^{(n-1)}\lambda^{(n)}} P_{\lambda^{(n)}}\Big].
\end{aligned}\end{equation}
Expanding $e^{p_1}$ in two ways, as demonstrated in the previous two sections, gives
\begin{equation}
	\sum_{\lambda^{(1)}\dots\lambda^{(n-1)};\lambda^{(n)}=\lambda} B_{\phi\lambda^{(1)}}B_{\lambda^{(1)}\lambda^{(2)}}\dots B_{\lambda^{(n-1)}\lambda^{(n)}} = \frac{ P_\lambda\{\beta^{-1}\delta_{k,1}\}}{\langle P_\lambda,P_\lambda\rangle},
\end{equation}
while the products in the first and second lines of \eqref{Wm1n} can be evaluated using \eqref{prodcells}. Altogether, we find
\begin{equation}
	Z_\beta = e^{W_{-1}/a_1}\cdot1 = 	 \sum_\lambda \frac{P_\lambda\{p_k\}}{\langle P_\lambda,P_\lambda\rangle}\frac{P_\lambda\{N\}P_\lambda\{N+\beta^{-1}\nu + \beta^{-1}-1\}}{P_\lambda\{a_1\beta^{-1}\delta_{k,1}\}},
\end{equation}
so that, in this theory, the correlators are
\begin{equation}
	\big\langle P_\lambda(z_1,\dots,z_N) \big\rangle=\frac{P_\lambda\{N\}P_\lambda\{N+\beta^{-1}\nu + \beta^{-1}-1\}}{P_\lambda\{a_1\beta^{-1}\delta_{k,1}\}}.
\end{equation}
This relation was conjectured and verified for $|\lambda|\leq9$ in \cite{CLZ}, and here we provided its complete proof. In particular, when $\beta=1$, for a  complex linear model $\int d^2Z e^{-\Tr ZZ^\dagger}$, where $Z$ is of size $N_1\times N_2$, one finds

\begin{equation}
	\big\langle s_\lambda(ZZ^\dagger) \big\rangle=\frac{s_\lambda\{N_1\}s_\lambda\{N_2 \}}{s_\lambda\{\delta_{k,1}\}}.
\end{equation}


\section*{Acknowledgments}

We thank Luca Cassia and Maxim Zabzine for helpful discussions. It is a pleasure to acknowledge the package \cite{RobertCoquereaux}, which helped verify and guide our proofs. This work has been supported by the TEAM programme of the Foundation for Polish Science co-financed by the European Union under the European Regional Development Fund (POIR.04.04.00-00-5C55/17-00).





\newpage

\bibliographystyle{JHEP}
\bibliography{superint_bib}

\end{document}